\documentclass[a4paper]{article}

\usepackage{cite}
\usepackage{amsmath,amssymb,amsfonts}
\usepackage{algorithm}
\usepackage{algorithmic}
\usepackage{comment}
\usepackage{graphicx}
\usepackage{textcomp}
\usepackage{xcolor}
\usepackage{listings}
\usepackage{balance}
\usepackage{authblk}

\newcommand{\CommentLine}[1]{
    \STATE  \textcolor{blue}{\textit{// #1  } }
}
\newcommand{\CommentInline}[1]{
      \hfill \textcolor{blue}{\textit{// #1 }} 
}

\providecommand{\keywords}[1]
{
  \small	
  \textbf{Keywords---} #1
}

\title{Towards Management of Energy Consumption in HPC Systems with Fault Tolerance

}

\author[1]{Marina Morán}
\author[1]{Javier Balladini}
\author[2]{Dolores Rexachs}
\author[3]{Enzo Rucci}

\affil[1]{Ing. de Computadoras - Fac. de Informática\\
Universidad Nacional del Comahue, Neuquén, Argentina\\
\authorcr
\{marina, javier.balladini\}@fi.uncoma.edu.ar}

\affil[2]{CAOS - Computer Architecture and Operating Systems\\
Universidad Aut\'onoma de Barcelona, Barcelona, Spain\\
\authorcr
dolores.rexachs@uab.es}

\affil[3]{III-LIDI, Facultad de Inform\'atica,
Universidad Nacional de La Plata - CIC\\
La Plata, Buenos Aires, Argentina\\
 \authorcr
erucci@lidi.info.unlp.edu.ar}

\date{{August 17, 2021}}

\begin{document}

\maketitle 

\begin{center}
\texttt{This is the accepted version of the manuscript that was sent to review to \textit{2020 IEEE Biennial Congress of Argentina (ARGENCON)} (ISBN 978-1-7281-5957-7\/20). This manuscript was finally accepted for publication on November 14th, 2020 and its final published version is available at https://doi.org/10.1109/ARGENCON49523.2020.9505498}
\end{center}

\begin{center}
\texttt{\textregistered 2021 IEEE. Personal use of this material is permitted.  Permission from IEEE must be obtained for all other uses, in any current or future media, including reprinting/republishing this material for advertising or promotional purposes, creating new collective works, for resale or redistribution to servers or lists, or reuse of any copyrighted component of this work in other works.}
\end{center}

\clearpage

\begin{abstract}

High-performance computing continues to increase its computing power and energy efficiency. However, energy consumption continues to rise and finding ways to limit and/or decrease it is a crucial point in current research. For high-performance MPI applications, there are rollback recovery based fault tolerance methods, such as uncoordinated checkpoints. These methods allow only some processes to go back in the face of failure, while the rest of the processes continue to run. In this article, we focus on the processes that continue execution, and propose a series of strategies to manage energy consumption when a failure occurs and uncoordinated checkpoints are used. We present an energy model to evaluate strategies and through simulation we analyze the behavior of an application under different configurations and failure time. As a result, we show the feasibility of improving energy efficiency in HPC systems in the presence of a failure.

\end{abstract}


\section{Introduction}
High Performance Computing (HPC) continues to increase its computing power and energy efficiency \cite{subramaniam2013trends, gao2016survey}. For example, the supercomputer Fugaku\footnote{https://www.top500.org/system/179807/}, number 1 of the Top500, presents 415.5 PFlop/s against the 148.6 PFlop/s of its predecessor, Summit. At the same time, these supercomputers are in the top ten of the Green500, with around 14.7 GFlops/W. However, energy consumption continues to increase; while the Summit computer has a consumption of 10 MW, Fugaku goes up to 28 MW. As this increase in energy consumption is not sustainable, it is necessary to reduce it.

A parallel message passing application can be affected by failures from computer system components. In this work, we refer to permanent failures, which cause fail-stops in MPI (Message Passing Interface), in homogeneous clusters. There are methods to continue with the execution in the presence of a failure. One of the most widely used methods is rollback-recovery through the use of checkpoints. When a node fails, it is possible to use an uncoordinated checkpoint scheme where the processes of the nodes that have not failed continue their execution. These processes will eventually stop when they need to communicate with a process that is recovering. As there will be a wait, we think that this scenario presents opportunities for energy savings. If waiting is unavoidable, what is the best strategy to consume less energy at that time? The challenge is to investigate what possibilities exist when one or more processes stop their execution. How to take advantage, from an energy-saving point of view, of the great benefit that uncoordinated checkpoints present by avoiding that all the application processes have to go back in the presence of a failure?

In this work, we evaluate a series of strategies that can be applied to improve energy efficiency when a failure occur. The strategies use the Advanced Configuration and Power Interface (ACPI), in particular we consider the use of Dynamic Voltage and Frequency Scaling (DVFS) techniques and system hibernation at the node level. By having a characterization of the energy consumption required to execute the application and its communication pattern, we estimate the execution and waiting times of the processes that do not fail. Then, by using a simulator that we have designed and developed, we can evaluate the use of the strategies. Using a simulator allows us to simplify a real system, reduce costs, and focus on essential features. In our case, it also allows us to have a flexible environment to experiment with different configurations. Also, we use a tool to present the results in a visually. 

Our objective is to know and manage the energy consumption of an HPC system, applying different strategies depending on the state of the application and the characteristics of the machine, when a permanent failure occurs and local rollback recovery is used, without increasing execution time. We create a model that allows predicting the energy consumption of an HPC application with its fault tolerance (FT) method under certain system conditions. The simulations show that in an interval of around 40 minutes it is possible to achieve around 70\% of energy saving. The main contributions of this work are:
\begin{itemize}
    \item The definition of a series of strategies for energy saving when a failure occurs and only failed processes have to rollback. 
    \item The design and development of a simulator oriented to evaluate the proposed strategies and to select the most convenient one from the energy point of view.
    \item The definition of a model that allows us to asses the impact of the strategies application on energy consumption under different scenarios.
\end{itemize}

The rest of the paper is organized as follows. Section \ref{sec:back_y_related} presents some preliminary concepts used in the article and some related works. Section \ref{sec:propuesta} describes the proposal and its motivation, and presents the energy model. The simulator and the experimental results are presented in section \ref{sec:experimentacion}. Finally, conclusions are summarized in section \ref{sec:conclusiones}.

\section{Background and Related Work}\label{sec:back_y_related}
The following subsections introduce some concepts that are used in the article, such as active waits in MPI, rollback recovery mechanism, and the states defined by ACPI. The last subsection presents some related works.

\subsection{Waits in MPI}
In MPI parallel applications, it may happen that one process must wait for another to send or receive a message. During these waits, the process can keep the processor busy by active-waiting, or release it, and use polling or interrupts. An active wait keeps the processor busy and consuming energy, without doing useful work. An idle wait can impact on application performance, due to C states transitions \cite{cesarini2018countdown}, among others. Various MPI implementations provide active wait as the default operating mode. As this operating mode is configurable, in this work we consider both cases.

\subsection{Rollback recovery}

A consistent global state can be found during a successful and fault-free execution of parallel computing. Inconsistent states occur because of failures. A fundamental goal of any rollback-recovery protocol is to lead the system to a consistent state after a failure. This method consists of periodically saving the state of the application in stable storage, which is known as a \textit{checkpoint}. At failure time, it is possible to restart the application from the last successfully saved state, which is called \textit{restart}. There are two main approaches: coordinated and uncoordinated checkpoints. In the case of coordinated checkpoints, the consistent global state is obtained by synchronizing all the processes at checkpoint time, and when a process fails, all processes restart from the last checkpoint. As we can see, all the application processes re-executing produce energy and performance overhead. In the case of uncoordinated checkpoints, processes take their checkpoints independently, avoiding synchronization overheads and I/O contention~\cite{levy2014using}. At failure time, only failed processes restart from the last checkpoint, using fewer resources for its recovery than a coordinated checkpoint. However, ensuring a consistent global state is not as straightforward as in the case of coordinated checkpoints. When a process restarts, orphaned and/or lost messages can appear, causing other processes to roll back to ensure consistency. This is called \textit{domino effect}, and there are different techniques to control it, such as the use of message logging~\cite{meyer2017hybrid}.

There are hybrid approaches to take advantage of coordinated and uncoordinated checkpoints. In this scheme, the processes are divided into groups. Within each group coordinated checkpoint is used, but between groups, uncoordinated checkpoint is used. There are different criteria for defining groups. For example, all processes running on the same node could be in a group, because when a node fails all its processes must restart \cite{castro2015fault}. Another way to define groups can be with processes that communicate frequently \cite{ho2008scalable}. The first approach is the one used in the present work.

\subsection{ACPI}\label{sub:backgroundACPI}
The ACPI specification provides an open standard that allows the operating system to manage the power of the computing system and provides advanced mechanisms for energy management\footnote{https://www.uefi.org/specifications}. The specification defines a series of global states and substates for the system. In the global \textit{sleeping} state G1 the computer consumes a small amount of power and applications are not executed. As the context is saved, the operating system does not need to restart when waking up. Latency for returning to the working state varies on the type of sleeping substates selected (S1-S4). In this work, we use performance and sleeping states (P and S states).

\subsection{Related Work}
Some works that take advantage of the waits of processes that do not rollback when a failure occurs. In \cite{bouteiller2013multi}, they look to improve the efficiency of the computer system by replacing the application when the waits are long enough. That way, while the failed processes are recovering, another application is allowed to advance. Most articles seek to regulate power consumption using DVFS. In \cite{knobloch2012determine} they analyze the active waits of an MPI application and evaluate potential energy savings by changing the clock frequency during those waits. Other works slow down the non-critical path to consume less power without substantially increasing execution time \cite{bhalachandra2017adaptive, hajiamini2019dynamic}. 
\cite{dichev2018energy} is the most similar proposal to this work, since they propose a localized rollback based on the data flow, and reduce the clock frequency of the waiting processes to the minimum possible. We evaluate other strategies, in addition to changing to the minimum frequency, and we do so both for the computation and waits of the processes that continue to execute.

\section{Proposal}\label{sec:propuesta}
In the following subsections, the motivation is reinforced, the strategies and their application are defined, and an energy model that allows estimating the energy savings achieved is shown.

\subsection{Motivation}
Uncoordinated checkpoints allow only the processes of the failed node to be restarted, while the others continue execution. After a failure, re-execution time depends on the time of the last checkpoint. The further from the last checkpoint the failure occurs, the longer will be the re-execution time. The duration of this re-execution time will affect the duration of the waits suffered by the processes that continue execution. During these waits, the processes consume energy without doing useful work. In an application with several processes, as is the case of HPC applications, this can mean a significant waste of resources. Managing these waits in order to achieve lower energy consumption motivates the present work.

\subsection{Strategies definition and application}\label{sub:politicas}
\begin{figure*}
  \centering
  \includegraphics[width=13cm]{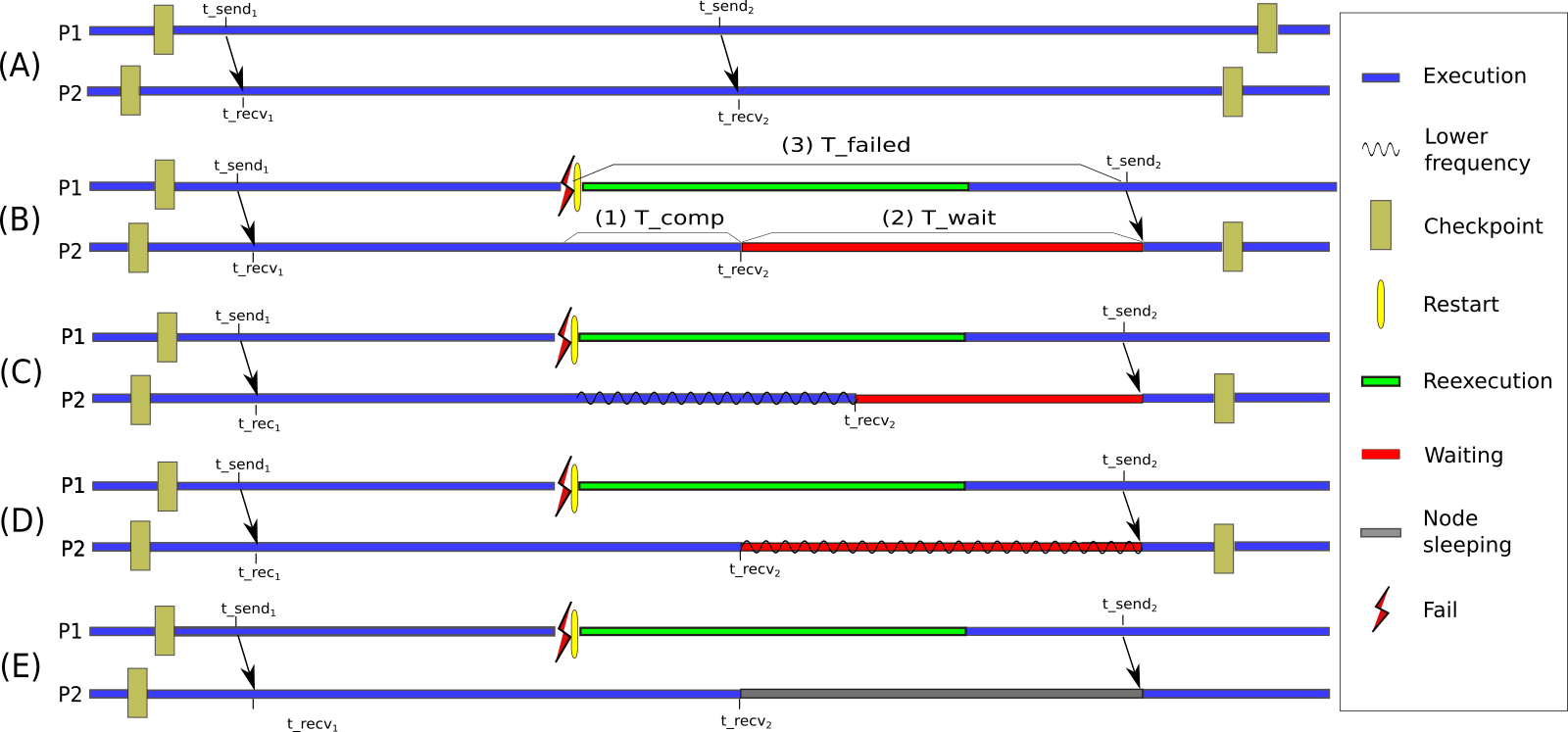}
  \caption{Application cases. (A) No failure. (B) Failure and no action. (C) Frequency change for the computational phase. (D) Frequency change in the waiting phase. (E) Sleeping in the waiting phase.}
  \label{fig:casos}
\end{figure*}

When a node fails, the strategy to be applied to surviving nodes is evaluated, analyzing the state of its processes. In order to improve energy efficiency and taking into account the impact on execution time, the strategies combination that should be applied to nodes that continue to execute after a failure is evaluated. Fig. \ref{fig:casos} shows different scenarios, where two processes, P1 and P2, are running on different nodes. Vertical rectangles indicate that the process is performing checkpoints. As uncoordinated checkpoints are used, they may be performed at different times, as shown in the figure. P1 sends two messages to P2, indicated by $t\_send_1$ and $t\_send_2$. The computation and wait phase of surviving process P2, form the \textit{intervention interval} and are indicated with 1 and 2 in case B. The computation phase comprises the execution of the application from the moment of failure until it is blocked waiting for communication with a process that has failed, at which point the waiting phase begins. It may happen that during the computation phase additional waits appear, but caused by other surviving processes and not by a process in recovery. It may also happen that during this phase checkpoints are perform. 

At failure time, which clock frequency is more convenient to use for the computation phase in combination with the action for the waiting phase is determined. The action for the waiting phase can be to sleep the node (in some state S1-S4, subsection \ref{sub:backgroundACPI}) or to switch to the minimum clock frequency. The strategies can be summarized as follows:

\begin{itemize}
    \item Frequency change for the computational phase (case C). 
    \item Frequency change for the waiting phase (case D).
    \item Sleeping for the waiting phase (case E).
\end{itemize}

\textbf{Case A} shows a failure-free execution, where messages are sent and received in a synchronized way as is in the application. \textbf{Case B} shows an execution where process P1 fails and must recover. After the failure, the restart is executed and the re-execution follows, indicated in green. We can see how the sending of the second message is delayed due to the failure, and the process P2 must wait, indicated in red. This is the case that will serve as a reference for the evaluation of the strategies. In the following cases, we observe the implementation of the strategies. In \textbf{case C}, the P2 node changes to a lower clock frequency during the computation phase, indicated by the wavy line. This makes P2 submit the second reception later, shortening the waiting phase. In \textbf{case D}, it was decided to change the clock frequency during the waiting phase. This action does not affect the waiting phase duration, but it does impact energy consumption. In \textbf{case E}, the P2 node goes to sleep during the waiting phase, reducing power consumption. These strategies can be applied together. For example, it could be the case that the clock frequency is changed for the computation phase and the waiting phase. As we can see, there are several possible scenarios where different actions must be evaluated and managed.
 
The evaluation of the strategies for the computing and waiting phase is done altogether. The selected configuration will be the one that achieves the lowest energy consumption for the intervention interval without affecting the execution time. For this, the frequency selected for the computing phase should not slow down the intervened process to the point that the recovered process must wait for it. Regarding the waiting phase, if the duration of this phase is long enough to get the node to sleep and to wake up, and at the same time lower energy consumption is achieved, then this option is selected. Otherwise, if the waits are configured to be active waits, the minimum frequency is selected, and if they are configured to be idle waits, then no action is taken.

Algorithm ~\ref{alg:strat-eval} shows the pseudocode for the evaluation of the strategies\footnote{See nomenclature in Table \ref{tab:parametros_modelo} Section \ref{sub:modeloEnergetico}}. This is executed at the time of failure for each surviving process. The first function, $estimate\_times(p)$, estimate $T\_comp$ and $T\_failed$, for process $p$ at maximum frequency. $T\_comp$ is the computational phase duration, and $T\_failed$ (indicated with 3 in case B of Fig. \ref{fig:casos}) is the time that elapses from the failure, until the recovered process is blocked by communication with the process $p$. The outputs of the algorithm are the frequency selected for the computational phase and the action to be taken for the waiting phase for each evaluated process.

\begin{algorithm}[ht!]
\caption{Pseudo-code for strategies evaluation}\label{alg:strat-eval}
	\begin{algorithmic}[1]
    \STATE estimate\_times(p); \CommentInline { estimate T\_comp and T\_failed.}
        \STATE min\_energy = MAX\_ENERGY; 
        \FOR {each frequency \textit{f}}
            \STATE comp\_time = T\_comp $\times \beta$(f) + N\_ckpt$\times$T\_ckpt$\times\gamma$(f);              \CommentInline { estimate the duration of the compute phase.  }
            \IF {comp\_time $>$ T\_failed} 
                 \CommentLine{the restarted process would have to wait.} 
                \PRINT "Frequency not allowed."
            \ELSE
             \CommentLine{ strategies are evaluated }
                \STATE wait\_time = T\_failed - T\_comp $\times \beta$(f); 
                \IF {active\_wait\_used}
                    \CommentLine{ estimate energy when using active waits.}
                    \STATE E\_awake\_wait = wait\_time $\times$ P\_active\_wait;
                \ELSE
                     \CommentLine{ estimate energy when using idle waits. }
                    \STATE E\_awake\_wait = wait\_time $\times$ P\_idle\_wait;
                \ENDIF
                \CommentLine{ evaluate if it is worth sleeping the node: wait is long enough and the energy saving is significant. }
                \STATE T\_sw =  T\_go\_sleep + T\_wakeup;
                \IF {(wait\_time $> \mu_1 \times$ T\_sw) \AND
                    (EI\_sleep\_wait $< \mu_2 \times$ E\_awake\_wait)}   
                   \CommentLine{ node goes to sleep. }
                    \STATE action\_wait\_time = \textit{sleep};
                    \STATE wait\_energy = E\_sw;
                \ELSE
                   \CommentLine{ node stays awake. }
                    \STATE action\_wait\_time = \textit{awake};
                    \STATE wait\_energy = E\_awake\_wait;
                \ENDIF
                \STATE compute\_energy = comp\_time $\times$ P(f);
                \STATE total\_energy = compute\_energy + wait\_energy;
                \IF {total\_energy $<$ min\_energy} \CommentLine{ save the strategy that 
          minimizes energy. }
                    \STATE compute\_frequency\_selected = f;
                    \STATE action\_wait\_time\_selected = 
                        action\_wait\_time;
                \ENDIF
            \ENDIF
        \ENDFOR
	\end{algorithmic}
\end{algorithm}

\subsection{Energy model}\label{sub:modeloEnergetico}
The model presented below estimates the energy savings achieved when the selected strategies are applied after a failure. The input data of the model is the selected strategy in combination with system, application and fault tolerance characterization data as indicated in Table \ref{tab:entradas_modelo}. Power and time can be obtained from characterizations as in \cite {moran2019prediction}. The application communication pattern can be obtained from the execution trace. Downtime refers to the time that elapses from the failure time until the restart begins. The clock frequency selected for the computational phase is $fb$, and $fa$ is the maximum available frequency. Table \ref{tab:parametros_modelo} details model parameters for reference. Modeled communications are blocking communications. Case B of Fig. \ref{fig:casos} reflects the situation where no strategies are applied, and serves as reference.

\begin{table}
\centering
\caption{Energy model inputs}
\begin{tabular}{ |p{3cm}|p{8cm}|  }
\hline
System data &  Power and time required to sleep and to wake up a node. \\
\hline
Application data & Power dissipated and slowdown of each frequency during the computation. Pattern and frequency of communication among processes. \\
\hline
Fault tolerance data & Checkpoint and restart duration. Power dissipated and slowdown of each frequency during checkpoint\\
\hline
Variables &  Checkpoint interval, failure time, downtime.\\
\hline
\end{tabular}
\label{tab:entradas_modelo}
\end{table}

The energy saving is estimated for each non-failed node $i$, and is the difference between the energy consumption with and without the use of the strategies, as shown in (\ref{eq:ahorro}). 

\begin{equation}
Energy\_saving_i(fa, fb) = ENI_i(fa) - EI_i(fb)
\label{eq:ahorro}
\end{equation}

We call the energy consumed by node $i$ without intervention $ENI_i$ (\underline{N}o \underline{I}ntervention, equation \ref{eq:ENI}),  while $EI_i$ (equation \ref{eq:EI}), is the Energy with Intervention. In these equations, energy estimation during the waiting phase varies whether or not active wait is used, as this affects the power dissipated. The equation for the energy of the computational phase is shown in (\ref{eq:Ecomp}), and for the waiting phases in the (\ref{eq:E_SleepWait}), (\ref{eq:E_active_wait}) and (\ref{eq:E_idle_wait}). Whenever we mention the energy consumed by a node, with or without intervention, we refer to the energy consumed during the \textit{intervention interval}. This interval is different for each node, starting at the failure and ending when some process of the node communicates with a recovered process.

To calculate the energy consumption during an interval of time we need to know its duration and the associated average power dissipation. Power is obtained from the model data input, while the time (equations \ref{eq:tcomp} and \ref{eq:T_wait)}) depends on two variables: the failure time and the strategy adopted. The failure time determines the value of $\alpha_{ij}$, which is the communication interval percentage that a process from node $i$ still needs to complete to reach the next communication point with a process from the failed node $j$. On the other hand, the adopted strategy will indicate us which is the most convenient clock frequency for the computation phase (\textit{fb}). The slowdown factor can be calculated from \textit{fb} and indicates how much slower the application runs with frequencies lower than the maximum one. The energy consumed to sleep and wake up the node will depend solely on the characteristics of the node and is a fixed value for each hibernation substate (S1-S4) of G1 implemented by the system. To simplify the equations we use a single value.

To sleep a node, two conditions must be met. First, the waiting time must be greater by a certain margin ($\mu_1$ in (\ref{eq:EIwait})), possibly zero, than the total time that the node requires to sleep and wake up. This is to prevent a recovered process from having to wait for an intervened process. Secondly, the energy consumption while sleeping, including the energy consumed while sleeping and waking up, must be lower by a certain margin ($\mu_2$ in (\ref{eq:EIwait})) than the consumption obtained if the node remains awake. The two options must be fulfilled to sleep the node; otherwise, the node remains awake. In this case, the consumption will be determined by the message waiting configuration. If the configuration indicates that active waits are used, the energy consumed is calculated using the power dissipated by the lowest available frequency. On the other hand, if idle waits are used, the processor is practically without work, and then the energy consumed is calculated using a power near to the base power.

\begin{equation}
ENI_i(fa)= E\_comp_i(fa) + E\_awake\_wait_i(fa)
\label{eq:ENI}
\end{equation}

\begin{equation}
EI_i(fb)= E\_comp_i(fb) + EI\_wait_i(fb)
\label{eq:EI}
\end{equation}

\begin{equation}
\begin{aligned}
E\_comp_{i}(f)= T\_comp_{i}(f) \times P\_comp(f) + \\
            N\_ckpt \times T\_ckpt(f) \times P\_ckpt(f)
\end{aligned}
\label{eq:Ecomp}
\end{equation}

\begin{equation}
T\_comp_{i}(f) = \alpha_{ij} \times I\_comm_{ij} \times \beta(f)
\label{eq:tcomp}
\end{equation}

\begin{equation}
T\_check(f) = check\_time \times \gamma(f)
\label{eq:tcheck}
\end{equation}

\begin{equation}
E\_awake\_wait_i(f) =
\begin{cases}
    E\_active\_wait_i(f) & \text{  if active wait}  \\
    E\_idle\_wait_i(f) & \text{  if idle wait}
\label{eq:E_awake_wait}
\end{cases}
\end{equation}

\begin{equation}
EI\_wait_i(fb) =
\begin{cases}
     EI\_sleep\_wait_i(fb)    & \\ 
        \quad\quad\text{if }  T\_wait_i(fb) > \mu_1 \times \\
        \quad\quad(T\_go\_sleep + T\_wakeup) \\
        \quad\quad\text{and } EI\_sleep\_wait_i(fb)  < \mu_2 \times \\
        \quad\quad E\_awake\_wait_i(fb)\\

    E\_active\_wait_i(fb) & \\ 
    \qquad\qquad\text{if active wait and no sleeping} \\

    E\_idle\_wait_i(fb) & \\ 
    \qquad\qquad\text{if idle wait and no sleeping}
\label{eq:EIwait}
\end{cases}
\end{equation}

\begin{equation}
\begin{aligned}
 EI\_sleep\_wait_i(fb)= & T\_go\_sleep \times P\_go\_sleep + \\
 & T\_sleep_i(fb) \times P\_sleep + \\ 
 & T\_wakeup \times P\_wakeup
\end{aligned}
\label{eq:E_SleepWait}
\end{equation}

\begin{equation}
E\_{active\_wait}_{i}(f) = T\_wait_{i}(f) \times P\_active\_wait
\label{eq:E_active_wait}
\end{equation}

\begin{equation}
E\_{idle\_wait}_{i}(f) = T\_wait_{i}(f) \times P\_idle\_wait
\label{eq:E_idle_wait}
\end{equation}

\begin{equation}
T\_sleep_i(f) = T\_wait_{i}(f) - T\_go\_sleep - T\_wakeup
\label{eq:T_sleep}
\end{equation}

\begin{equation}
T\_{wait}_{i}(f)= T\_failed_{i} - T\_comp_{i}(f)
\label{eq:T_wait)}
\end{equation}

\begin{equation}
T\_failed_{i} = T\_recover + \alpha_{ji} \times I\_comm_{ij}
\label{eq:T_failed)}
\end{equation}

\begin{equation}
T\_recover = T\_down + T\_restart + T\_reexec
\label{eq:T_recover)}
\end{equation}

\begin{table}[htbp]
\caption{Parameters}
\begin{center}
\begin{tabular}{|p{3cm}|p{10cm}|}
\hline
Parameter Name & Description \\
\hline
$E\_comp_i(f)$ & Energy consumed by node $i$ during the computing phase, at frequency $f$. \\
\hline
$E\_awake\_wait_i(f)$     & Energy consumed by node $i$ during the waiting phase when it remains awake.\\
\hline
$EI\_wait_i(f)$     & Energy consumed by node $i$ during the waiting phase when intervention takes place.\\
\hline
$E\_active\_wait_i(f)$ & Energy consumed by node $i$ during the active waiting phase at the minimum frequency.\\
\hline
$E\_idle\_wait_i(f)$ & Energy consumed by node $i$ during the waiting phase when using idle wait.\\
\hline
$EI\_sleep\_wait(f)$ & Energy consumed by node $i$ during the waiting phase when it goes to sleep. \\
\hline
$T\_ckpt(f)$ & Checkpoint duration running at frequency $f$.\\
\hline
$T\_comp_{i}(f)$ & Computation phase duration when node $i$ executes at the frequency $f$.\\
\hline
$T\_go\_sleep$ $T\_wakeup$ & Times required by a node to sleep and wake up, respectively.\\
\hline
$T\_{failed_{i}}$ & Time required by a failed process to recover and to block with a process of node $i$.\\
\hline
$T\_recover$ & Time required by a failed process to recover and return to the point where the failure occurred. \\
\hline
$T\_down$ & Downtime. \\
\hline
$T\_rest$ & Restart duration at maximum frequency. \\
\hline
$T\_reexec$ & re-execution time at maximum frequency. \\
\hline
$T\_sleep_i(f)$ & Time that node $i$ is sleeping (without considering the time to go to sleep and to wakeup).\\
\hline
$T\_wait_{i}(f)$ & Waiting phase duration of node $i$ process when compute phase was executed at frequency $f$. \\
\hline
$T\_ckpt$ & Checkpoint duration. \\
\hline
$P\_go\_sleep$ $P\_wakeup$ & Power dissipated while sleeping and waking up a node, respectively.  \\
\hline
$P\_sleep$ & Power dissipated when node is sleeping.  \\
\hline
$P\_comp(f)$ & Power dissipated by the application when running at frequency $f$.\\
\hline
$P\_ckpt(f)$ & Power dissipated during checkpoint at frequency $f$.\\
\hline
$P\_active\_wait$ & Power dissipated during and active wait. \\
\hline
$P\_idle\_wait$ & Power dissipated during and idle wait. \\
\hline
$N\_ckpt$ & Number of checkpoints in the intervention interval.\\
\hline
$\alpha_{ij}$ & Percentage of the communication interval between process $i$ and process $j$ that still remains to be executed for process $i$ to block in a communication with process $j$.\\
\hline
$I\_comm_{ij}$ & Duration of the communication interval between process $i$ and process $j$ when executing at maximum frequency.\\
\hline
$\beta(f)$ $\gamma(f)$ & Slowdown of instruction and checkpoint execution when frequency $f$ is used. \\
\hline
$\mu_1$ $\mu_2$ & Time and energy threshold to determine whether or not to sleep a node. \\
\hline
\end{tabular}
\label{tab:parametros_modelo}
\end{center}
\end{table}

\section{Experimentation and Results Analysis}\label{sec:experimentacion}
The following subsections describe the simulator, the simulations conditions, and the configuration data. After that, the results obtained are analyzed.

\subsection{Simulator}\label{simulacion}
We have developed an event-based simulator that uses the SMPL library written in C language \cite{macdougall1987simulating}. This simulator allows us to evaluate the strategies under different system configurations, different characteristics of the application and different failure times. The failure of a node in a parallel message passing application, with uncoordinated checkpoints at the system level is simulated. To simplify the first version of the simulator, a single process per node is simulated, the node's representative process. We call the representative process of the node to the process that first block due to communication with some process of the failed node. The strategy selected when evaluating the representative process is applied to the node. Checkpoints can be triggered by events or by time; as we seek to simulate transparent checkpoint to the application, we activate it by time. The moving ahead of checkpoints is simulated. If a process is going to block by communication with a recovering process, and its last checkpoint happened some time ago, the process performs a checkpoint before blocking. In this way, useless waiting time is used by a checkpoint. The checkpoint files are stored in a parallel file system external to the nodes. The message log and the domino effect have not been considered. The simulated MPI functions are blocking  $send$ synchronous mode ($MPI\_Ssend$) and blocking $receive$ ($MPI\_Recv$). The messages have a fixed size. The overhead caused by the strategies evaluation and implementation is not computed. Processes that indirectly block with a recovering process are not evaluated in this version of the simulator. For example, process A is waiting for the failed process B, and process C blocks with process A, then process C indirectly blocks with the recovering process B. The simulator inputs are the same as the energy model described in subsection \ref{sub:modeloEnergetico}, and are detailed in Table \ref{tab:entradas_modelo}.

At the time of failure, the simulator evaluates each surviving process with each of the clock frequencies provided, and determines the best strategy to apply. The simulator output includes the estimated energy savings when applying the selected strategy, and a trace to visualize the behavior of the application. The trace is visualized with the Paraver\footnote{http://www.bsc.es/computer-sciences/performance-tools/paraver} tool, a flexible HPC application performance analysis and visualization tool.

\subsection{Experimental settings}

Table \ref{tab:conf1} shows dissipated power and slowdown factor ($\beta$ and $\gamma$) obtained from measurements on a six-core Intel Xeon E5-2630 node, with a frequency range of 1.2 GHz to 2.8 GHz (with the mechanism Intel Turbo Boost disabled). The base power is 60W. The node sleep and wake times are set at 25 and 5 seconds respectively, and the average powers at 51 and 91 watts respectively. The average power dissipated while the node is sleeping is 12 watts. These values were obtained from \cite{xavier2017modeling} and correspond to S3 sleeping mode. The checkpoint duration is set to two minutes, and the MPI waits are configured as active waits, except otherwise indicated. The scenarios present four processes (or nodes) and the node that fails is the corresponding for process 0.

\begin{table}
\centering
\caption{Power and slowdown at different clock frequencies}
\resizebox{\columnwidth}{!}{%
\begin{tabular}{ |c|c|c|c|c|  }
\hline
 &  \multicolumn{2}{|c|}{Application} &  \multicolumn{2}{|c|}{Checkpoint}  \\
\hline
Frequency (GHz) & Average Power (W) & $\beta$ & Average Power (W) & $\gamma$ \\
\hline
2.8 & 166 & 1   & 150 & 1    \\
2.1 & 148 & 1.2 & 142 &  1.1 \\
1.7 & 139 & 1.5 & 131 &  1.2 \\
1.2 & 126 & 2.1 & 125 &  1.4 \\
\hline
\end{tabular}
}
\label{tab:conf1}
\end{table}

\subsection{Results analysis}
Different simulated scenarios are discussed in this subsection. For each scenario, the particular configuration data is indicated. For space reasons, the trace is shown for two representative cases. In these figures, frequency and S-state changes are indicated by flags, checkpoints are indicated in brown, downtime, restart and re-execution are indicated by light blue, yellow and green respectively. The thin lines are communications, waits are indicated in red, and sleeping node in gray. The strategies adopted and the estimated energy saved are summarized in Table \ref{tab:accionesEscenarios}. In this table, column N indicates the node number, the Action column indicates the strategy applied, column T indicates the phase duration, and column TT the total duration. The last columns show the savings in joules, joules per second, and percent.

\subsubsection*{\textbf{Scenario 1: Short re-execution time}}
\begin{figure*}
  \centering
  \includegraphics[width=\textwidth,height=0.20\textheight]{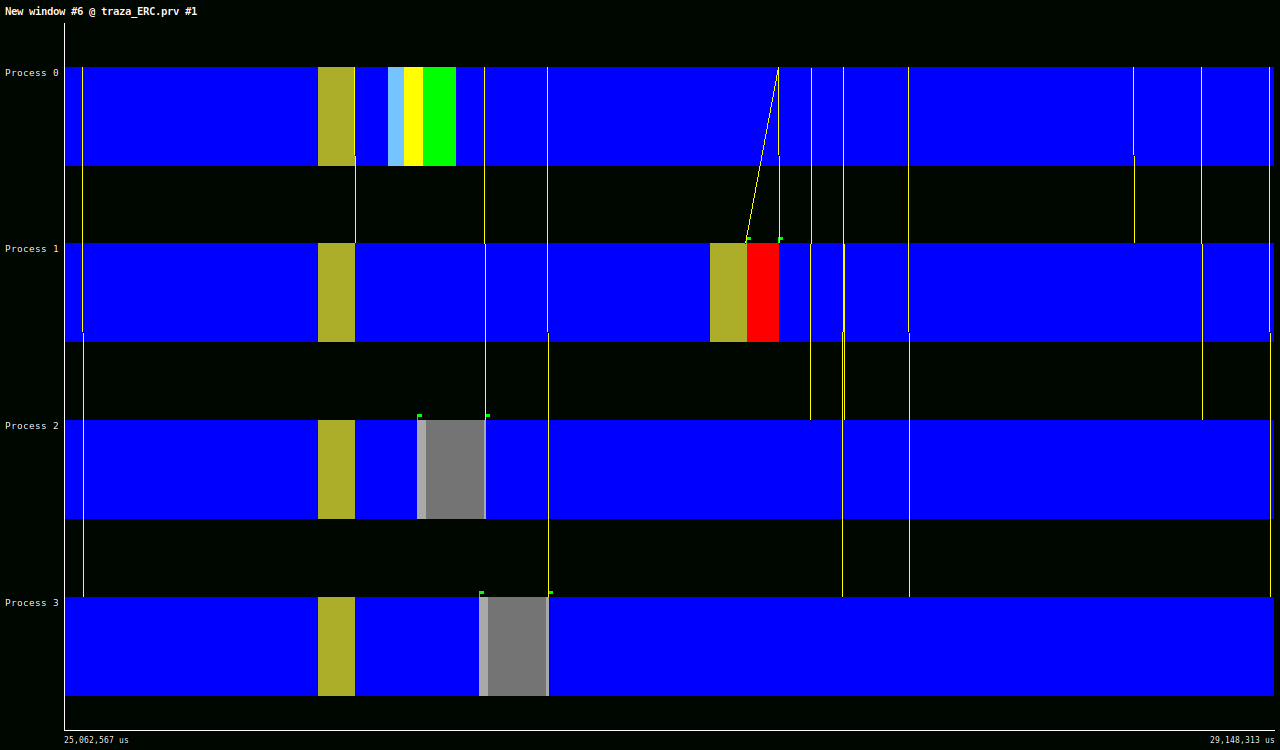}
  \caption{Scenario 1: Short re-execution time.}
  \label{E_reejec_corta}
\end{figure*}

The moment of failure is configured to occur immediately after a checkpoint. The visualization of this scenario can be seen in Fig. \ref{E_reejec_corta}. For the computation phase of process 1, the frequency is not changed because, with any of them (other than the maximal), the process would arrive late to the point of synchronization with the recovered process, and strategies that affect the execution time are not applied. For example, let us see what happens with the 2.1 GHz frequency. With this frequency, the computation phase would take approximately 21 minutes, while the recovered process will be waiting for process 1 in approximately 20 minutes. As for the waiting phase, as this scenario is configured with active waits, and the wait duration is long enough to justify a frequency change, it is changed to the minimum frequency. It can be observed that the checkpoint is moving ahead before starting the waiting phase (subsection \ref{simulacion}), reducing it. Nodes 2 and 3 do not change their frequency during the computation phase either, but unlike node 1, it is because they will go to sleep during the waiting phase. Therefore, it is not convenient to slow down the computation phase. With these actions, node 1 would achieve an energy saving of 2\% in an interval of 20 minutes, node 2 a saving of 60\% in an interval of 5 and a half minutes, and node 3 a saving of 50\% in a 7-minute interval. Nodes 2 and 3 achieve the highest energy savings, far from the savings obtained by node 1. If we see the j/s, node 1 achieves a 40 j/s saving, while node 2 and 3 achieve 148 j/s saving. This difference can be explained because nodes 2 and 3 sleep in the waiting phase.

\subsubsection*{\textbf{Scenario 2: Long re-execution time}}
\begin{figure*}
  \centering
  \includegraphics[width=\textwidth,height=0.20\textheight]{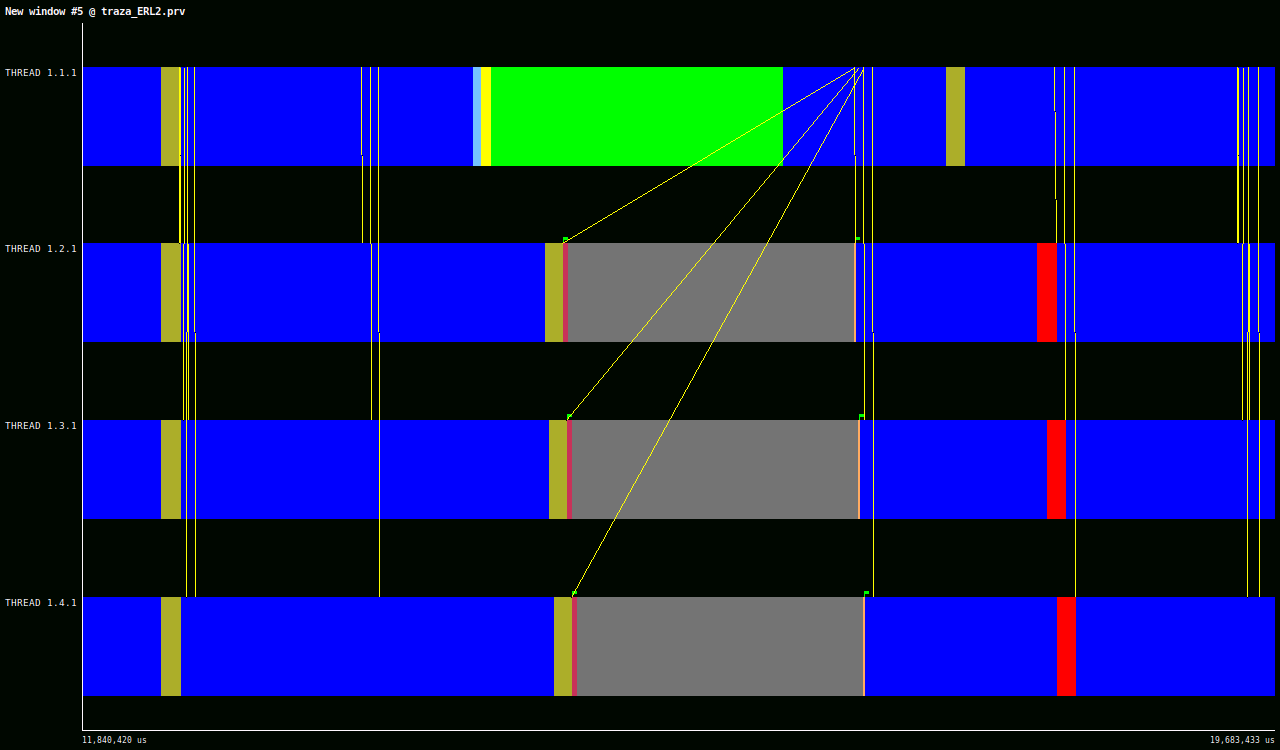}
  \caption{Scenario 2: Long re-execution time.}
  \label{fig:E_reejec_larga}
\end{figure*}

This scenario, which can be seen in Fig. \ref{fig:E_reejec_larga}, has configured the time of failure far from the checkpoint to evaluate what happens with a long re-execution time. In this scenario, checkpoints are moving ahead in all three processes. The waiting phases of the three surviving processes are very long, and the three nodes will go to sleep at this phase. For this reason, it is not advisable to change the frequency of the computation phase, because this would shorten the waiting phase, and therefore savings too. With these actions, the intervened nodes would be able to consume 70\% less energy during the intervention interval, which is around 42 minutes. This scenario achieves better results than the previous one due to its long waits where the nodes go to sleep.

\subsubsection*{\textbf{Scenario 3: Long re-execution time and frequency behaviour change}}
This scenario presents the same configuration as the previous one, but we now assume an application where the clock frequencies impact differently. We reflect this by changing the dissipated power and the slowdown (see Table \ref{tab:conf1}). If we decrease the dissipated power by 2W, and increase the slowdown by one tenth, the nodes change their frequency during the computation phase, unlike the previous scenario. During the waiting phase, all nodes sleep. With these actions, the intervened nodes would be able to consume 70\% less energy during the intervention interval, which is approximately 42 minutes. We can see that frequency change in the computational phase did not impact energy savings, which is almost the same as the savings in the previous scenario.

\subsubsection*{\textbf{Scenarios 4 and 5: Short re-execution time with and without active waits}}
In these scenarios, we have a short re-execution time, and active waiting is disabled for scenario 5. In both scenarios, the actions selected for the computation phase is to change the frequency, probably because the nodes are not going to sleep, since the waits are not long enough for it. In scenario 4 the frequency is changed to 1.2 Ghz in node 1 and to 1.7 Ghz in nodes 2 and 3, while in scenario 5 it is changed to 2.1 GHz in the three nodes. As indicated by the defined strategies, in scenario 4 (with active waiting), the waiting phase is changed to the minimum frequency, and in scenario 5 (with idle waiting) the waiting phase is left without action. With these actions, scenario 4 achieves energy savings between 17\% and 24\% compared to the savings of 0.17\% in scenario 5,  with both intervention intervals during the same time. This shows the impact on energy consumption of using active waits. Even when the scenario presents short waiting times, if the system is configured with active waiting, the use of the strategies achieves considerable energy savings.

\subsubsection*{\textbf{Scenario 6: Long re-execution time without moving ahead checkpoints}}
This scenario has the same configuration as the long re-execution scenario (scenario 2), but configured so that there is no moving ahead of checkpoints. Let us remember that in the reference scenario, the three processes move ahead the checkpoint before entering the waiting phase. The actions selected in this case are the same as those in scenario 2 (as expected) and the energy savings are slightly higher, around 74\%, compared to 70\% in the reference scenario. This difference can be explained by the aggregate consumption of the early checkpoint. 

These scenarios allowed us to observe that the strategies can achieve significant energy savings, especially when re-execution times are long. Additionally, we found that active waiting presents interesting opportunities for energy savings.

\begin{table}[]
\centering
\caption{Selected actions and energy savings for each node by scenario}
\resizebox{\columnwidth}{!}{%
\begin{tabular}{ |c|c|r|c|r|r|r|r|r| }
\hline
 &  \multicolumn{2}{|c|}{Compute phase} &  \multicolumn{2}{|c|}{Wait phase} & \multicolumn{4}{|c|}{}  \\
\hline
N & Action & T (m) & Action & T (m) & TT (m) & Save (J)   & Save (J/s) & Save (\%) \\
\hline
\multicolumn{9}{|c|}{Scenario 1: Short re-execution time} \\
\hline
1 & No action   & 18.20 & 1.2 GHz & 1.83  & 20.03  & 4,400.00   & 40.00      & 2.23      \\
2 & No action   & 1.73  & sleep  & 3.83  & 5.56   & 34,034.60  & 148.04     & 61.44     \\
3 & No action   & 3.23  & sleep  & 3.83  & 7.06   & 34,034.60  & 148.04     & 48.40     \\
\hline
\multicolumn{9}{|c|}{Scenario 2: Long re-execution time} \\
\hline
1 & No action  & 10.02 & sleep  & 32.00 & 42.02  & 294,294.60 & 153.29     & 70.64     \\
2 & No action  & 10.52 & sleep  & 32.00 & 42.52  & 294,294.60 & 153.29     & 69.81     \\
3 & No action  & 11.02 & sleep  & 32.00 & 43.02  & 294,294.60 & 153.29     & 69.00     \\
\hline
\multicolumn{9}{|c|}{Scenario 3: Long re-execution time and frequency behaviour change } \\
\hline
1 & 2.1 GHz   & 11.02  & sleep  & 31.20 & 42.02  & 291,346.88  & 115.56   & 70.75     \\
2 & 2.1 GHz   & 11.57  & sleep  & 30.95 & 42.52  & 291,448.88 & 114.24    & 69.94     \\
3 & 2.1 GHz   & 12.12  & sleep  & 30.90 & 43.02  & 291,550.88 & 112.96    & 69.15    \\
\hline
\multicolumn{9}{|c|}{Scenario 4: Short re-execution time with active waits} \\
\hline
1 & 1.2 GHz       & 4.93  & 1.2 GHz & 0.09  & 5.01   & 12,032.00  & 40.00      & 24.10     \\
2 & 1.7 GHz       & 4.15  & 1.2 GHz & 1.28  & 5.43   & 9,798.90   & 30.08      & 18.12     \\
3 & 1.7 GHz       & 4.77  & 1.2 GHz & 1.08  & 5.85   & 10,311.40  & 29.39      & 17.71     \\
\hline
\multicolumn{9}{|c|}{Scenario 5: Short re-execution time without active waits} \\
\hline
1 & 2.1GHz       & 2.82  & No action    & 2.20  & 5.01   & 56.32      & 0.33       & 0.17      \\
2 & 2.1GHz       & 3.32  & No action   & 2.11  & 5.43   & 66.32      & 0.33       & 0.18      \\
3 & 2.1GHz       & 3.82  & No action    & 2.03  & 5.85   & 76.32      & 0.33       & 0.18      \\
\hline
\multicolumn{9}{|c|}{Scenario 6: Long re-execution time without moving ahead checkpoints} \\
\hline
1 & No action          & 8.02  & sleep  & 34.00 & 42.02  & 312,774.60 & 153.33     & 74.74     \\
2 & No action          & 8.52  & sleep  & 34.00 & 42.52  & 312,774.60 & 153.33     & 73.86     \\
3 & No action          & 9.02  & sleep  & 34.00 & 43.02  & 312,774.60 & 153.33     & 73.00    \\
\hline
\end{tabular}
}
\label{tab:accionesEscenarios}
\end{table}

\section{Conclusions and Future Work}\label{sec:conclusiones}
Energy saving opportunities exists in a rollback recovery scheme where only some processes must go back and reexecute. We have proposed and analyzed different strategies to apply to the nodes of the surviving processes and we presented a model that allows estimating the energy savings achieved by applying these strategies. By using a simulator we showed the behavior of an application under different configurations and failure times. The simulations showed the validity of the proposed strategies to achieve significant energy savings and, in all analyzed cases, these saving were achieved without increasing the application execution time. In an interval of around 40 minutes it was possible to achieve energy savings of around 70\%. In this way, we showed the feasibility of improving energy efficiency in HPC systems in the presence of a failure.

Among the future works, we plan to continue the development of the simulator. The main extension, which can lead to greater energy savings, is to include the processes that indirectly block with a failed process in the evaluation and application of strategies. Additionally, implementing the strategies on a real system to verify the obtained results is an open line of investigation.

\section*{Acknowledgments}

This research has been supported by the Agencia Estatal de
Investigación (AEI), Spain and the Fondo Europeo de Desarrollo
Regional (FEDER) UE, under contract TIN2017-84875-P and partially
funded by a research collaboration agreement with the Fundacion
Escuelas Universitarias Gimbernat (EUG).

\balance

\bibliographystyle{IEEEtran}
\bibliography{citas.bib}

\begin{thebibliography}{10}
\providecommand{\url}[1]{#1}
\csname url@samestyle\endcsname
\providecommand{\newblock}{\relax}
\providecommand{\bibinfo}[2]{#2}
\providecommand{\BIBentrySTDinterwordspacing}{\spaceskip=0pt\relax}
\providecommand{\BIBentryALTinterwordstretchfactor}{4}
\providecommand{\BIBentryALTinterwordspacing}{\spaceskip=\fontdimen2\font plus
\BIBentryALTinterwordstretchfactor\fontdimen3\font minus
  \fontdimen4\font\relax}
\providecommand{\BIBforeignlanguage}[2]{{%
\expandafter\ifx\csname l@#1\endcsname\relax
\typeout{** WARNING: IEEEtran.bst: No hyphenation pattern has been}%
\typeout{** loaded for the language `#1'. Using the pattern for}%
\typeout{** the default language instead.}%
\else
\language=\csname l@#1\endcsname
\fi
#2}}
\providecommand{\BIBdecl}{\relax}
\BIBdecl

\bibitem{subramaniam2013trends}
B.~Subramaniam, W.~Saunders, T.~Scogland, and W.-c. Feng, ``Trends in
  energy-efficient computing: A perspective from the green500,'' in \emph{2013
  International Green Computing Conference Proceedings}.\hskip 1em plus 0.5em
  minus 0.4em\relax IEEE, 2013, pp. 1--8.

\bibitem{gao2016survey}
Y.~Gao and P.~Zhang, ``A survey of homogeneous and heterogeneous system
  architectures in high performance computing,'' in \emph{2016 IEEE
  International Conference on Smart Cloud (SmartCloud)}.\hskip 1em plus 0.5em
  minus 0.4em\relax IEEE, 2016, pp. 170--175.

\bibitem{cesarini2018countdown}
D.~Cesarini, A.~Bartolini, P.~Bonf{\`a}, C.~Cavazzoni, and L.~Benini,
  ``Countdown: A run-time library for application-agnostic energy saving in
  {MPI} communication primitives,'' in \emph{Proceedings of the 2nd Workshop on
  AutotuniNg and aDaptivity AppRoaches for Energy efficient HPC Systems}, 2018,
  pp. 1--6.

\bibitem{levy2014using}
S.~Levy, B.~Topp, K.~B. Ferreira, D.~Arnold, P.~Widener, and T.~Hoefler,
  ``Using simulation to evaluate the performance of resilience strategies and
  process failures,'' \emph{Sandia Labs, Tech. Report SAND2014-0688}, 2014.

\bibitem{meyer2017hybrid}
H.~Meyer, R.~Muresano, M.~Castro-Le{\'o}n, D.~Rexachs, and E.~Luque, ``Hybrid
  message pessimistic logging. improving current pessimistic message logging
  protocols,'' \emph{Journal of Parallel and Distributed Computing}, vol. 104,
  pp. 206--222, 2017.

\bibitem{castro2015fault}
M.~Castro-Le{\'o}n, H.~Meyer, D.~Rexachs, and E.~Luque, ``Fault tolerance at
  system level based on {RADIC} architecture,'' \emph{Journal of Parallel and
  Distributed Computing}, vol.~86, pp. 98--111, 2015.

\bibitem{ho2008scalable}
J.~C. Ho, C.-L. Wang, and F.~C. Lau, ``Scalable group-based checkpoint/restart
  for large-scale message-passing systems,'' in \emph{2008 IEEE International
  Symposium on Parallel and Distributed Processing}.\hskip 1em plus 0.5em minus
  0.4em\relax IEEE, 2008, pp. 1--12.

\bibitem{bouteiller2013multi}
A.~Bouteiller, F.~Cappello, J.~Dongarra, A.~Guermouche, T.~H{\'e}rault, and
  Y.~Robert, ``Multi-criteria checkpointing strategies: Response-time versus
  resource utilization,'' in \emph{European Conference on Parallel
  Processing}.\hskip 1em plus 0.5em minus 0.4em\relax Springer, 2013, pp.
  420--431.

\bibitem{knobloch2012determine}
M.~Knobloch, B.~Mohr, and T.~Minartz, ``Determine energy-saving potential in
  wait-states of large-scale parallel programs,'' \emph{Computer
  science-research and development}, vol.~27, no.~4, pp. 255--263, 2012.

\bibitem{bhalachandra2017adaptive}
S.~Bhalachandra, A.~Porterfield, S.~L. Olivier, and J.~F. Prins, ``An adaptive
  core-specific runtime for energy efficiency,'' in \emph{IEEE International
  Parallel and Distributed Processing Symposium}, 2017, pp. 947--956.

\bibitem{hajiamini2019dynamic}
S.~Hajiamini, B.~Shirazi, A.~Crandall, and H.~Ghasemzadeh, ``A dynamic
  programming framework for {DVFS}-based energy-efficiency in multicore
  systems,'' \emph{IEEE Transactions on Sustainable Computing}, vol.~5, no.~1,
  pp. 1--12, 2019.

\bibitem{dichev2018energy}
K.~Dichev, K.~Cameron, and D.~S. Nikolopoulos, ``Energy-efficient localised
  rollback via data flow analysis and frequency scaling,'' in \emph{Proceedings
  of the 25th European MPI Users' Group Meeting}, 2018, pp. 1--11.

\bibitem{moran2019prediction}
M.~Mor{\'a}n, J.~Balladini, D.~Rexachs, and E.~Luque, ``Prediction of energy
  consumption by checkpoint/restart in {HPC},'' \emph{IEEE Access}, vol.~7, pp.
  71\,791--71\,803, 2019.

\bibitem{macdougall1987simulating}
M.~H. MacDougall, \emph{Simulating computer systems: techniques and
  tools}.\hskip 1em plus 0.5em minus 0.4em\relax MIT press, 1987.

\bibitem{xavier2017modeling}
M.~G. Xavier, F.~D. Rossi, C.~A. De~Rose, R.~N. Calheiros, and D.~G. Gomes,
  ``Modeling and simulation of global and sleep states in {ACPI}-compliant
  energy-efficient cloud environments,'' \emph{Concurrency and Computation:
  Practice and Experience}, vol.~29, no.~4, p. e3839, 2017.

\end{thebibliography}

\end{document}